\documentclass[conference,10pt]{IEEEtran}

\usepackage{graphicx, subfigure}
\usepackage{url}
\usepackage{nicefrac}
\usepackage[latin1]{inputenc}
\usepackage[super, negative]{nth}
\usepackage{bytefield}
\usepackage{indentfirst}
\usepackage{bnf}
\graphicspath{{.}{./Pictures/}}

\newcommand{\dfn}[1]{\textit{#1}}

\title{Implementation and Deployment of a Distributed\\
Network Topology Discovery Algorithm}

\author{\authorblockN{Benoit Donnet\authorrefmark{1}\authorrefmark{2},
Bradley Huffaker\authorrefmark{2},
Timur Friedman\authorrefmark{1},
kc claffy\authorrefmark{2}}
\authorblockA{\authorrefmark{1}Université Pierre \& Marie Curie --
Laboratoire LiP6-CNRS, UMR 7606, Paris, France\\
email: \url{{benoit.donnet, timur.friedman}@lip6.fr}}
\authorblockA{\authorrefmark{2}CAIDA -- San Diego Supercomputer Center,
San Diego, USA\\
email: \url{{benoit, bhuffake, kc}@caida.org}}}

\begin{document}
\maketitle

\begin{abstract}
  In the past few years, the network measurement community has been 
  interested in the problem of internet topology discovery using a 
  large number (hundreds or thousands) of measurement monitors.  The 
  standard way to obtain information about the internet topology is 
  to use the traceroute tool from a small number of monitors.  Recent 
  papers have made the case that increasing the number of monitors 
  will give a more accurate view of the topology. However, scaling up 
  the number of monitors is not a trivial process. Duplication of 
  effort close to the monitors wastes time by reexploring well-known 
  parts of the network, and close to destinations might appear to be 
  a distributed denial-of-service (DDoS) attack as the probes 
  converge from a set of sources towards a given destination. In 
  prior work, authors of this report proposed Doubletree, an algorithm 
  for cooperative topology discovery, that reduces the load on the 
  network, i.e., router IP interfaces and end-hosts, while 
  discovering almost as many nodes and links as standard approaches 
  based on traceroute. This report presents our open-source and freely 
  downloadable implementation of Doubletree in a tool we call 
  traceroute@home. We describe the deployment and validation of 
  traceroute@home on the PlanetLab testbed and we report on the 
  lessons learned from this experience.   We discuss how 
  traceroute@home can be developed further and discuss ideas for 
  future improvements.
\end{abstract}

\section{Introduction}\label{intro}
%%%%%%%%%%%%%%%%%%%%%%
For some time, the problem of internet topology discovery has drawn 
the attention of the network measurement community.  One can see the 
internet topology at three different levels.  The first one, the 
\dfn{IP interface level}, considers internet protocol (IP) interfaces 
of routers and end systems. Usually, this topology is obtained by 
using data collected with the probing tool 
\dfn{traceroute}~\cite{traceroute}. Traceroute is a networking tool 
that allows one to discover IP interfaces along the path that data 
packets take to go from a \dfn{source} machine or \dfn{monitor} to a 
\dfn{destination} machine.  The second level, \dfn{the router level}, 
is an aggregation of the IP level.  It can be obtained by using a 
technique called \dfn{alias resolution}~\cite{heuristics, rocketfuel, 
onRoutes, pathDiversity}. The idea is to summarize all the IP 
addresses of a router into a single identifier.  Finally, the \dfn{AS 
level} provides information about the connectivity of autonomous 
systems (ASes). An AS is a set of routers that are under the same 
administrative control.  The ASes are interconnected in order to 
allow IP packets to transit from one network to another, so providing 
global connectivity.  The connectivity information might be 
inferred from BGP tables~\cite{routeview}, BGP being the routing 
protocol used between ASes.  Note that the special problem of
determining the topology within a single AS is a separate area of
inquiry.  It is not, strictly speaking, internet topology discovery,
and it is considerably helped by the privileged access available to
the administrator of an AS.

\begin{figure}[!t]
  \begin{center}
\includegraphics[width=7cm]{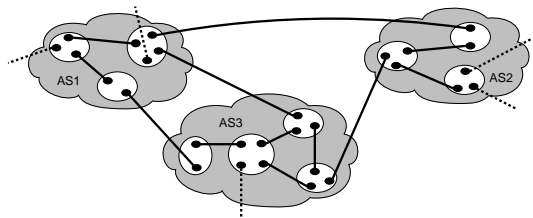}
  \end{center}
  \caption{The different levels of topology}
  \label{intro.topo}
\end{figure}
 
Fig.~\ref{intro.topo} illustrates the three levels of the internet 
topology.  Black dots represents router interfaces, blank shapes 
stand for routers and shaded areas for ASes.  The plain and dotted 
lines correspond to links.  The IP interface level is illustrated by 
the links between routers.  The router level is obtained when all 
interfaces of a router are grouped in a single identifier.  Finally, 
the AS level is obtained when we look only at ASes and the links 
between them.
 
This report focuses on the IP interface level.  More specially, it 
describes the implementation and deployment story of a topology 
discovery algorithm, \dfn{Doubletree}~\cite{DonnetSigmetrics2005}, 
that reduces load on the network, i.e., router IP interfaces and 
end-hosts, while discovering nearly the same set of nodes and links 
as standard approaches based on traceroute.  Doubletree was proposed 
by authors of this report.

Today's most extensive tracing system at the IP interface level, 
\dfn{skitter}~\cite{skitter}, uses 24 monitors, each targeting on the 
order of one million destinations.  Authors of this report are 
responsible for skitter.  In the fashion of skitter, 
\dfn{scamper}~\cite{scamper} makes use of several monitors to 
traceroute IPv6 networks. Other well known systems, such as \dfn{RIPE 
NCC TTM}~\cite{ripeNccTtm} and \dfn{NLANR AMP}~\cite{nlanrAmp}, each 
employ a larger set of monitors, on the order of one- to two-hundred, 
but they avoid probing outside their own network. However, recent 
work has indicated the need to increase the number of traceroute 
sources in order to obtain a more complete topology 
measurement~\cite{sampling, tracerouteSampling}.  Indeed, it has been 
shown that reliance upon a relatively small number of monitors to 
generate a graph of the internet can introduce unwanted biases. 

One way of rapidly creating a large distributed monitor 
infrastructure would be to deploy traceroute monitors in an easily 
downloadable and readily usable piece of software, such as a 
screensaver.  This was first proposed by J\"org Nonnenmacher, as 
reported by Cheswick et al.~\cite{mapping}.  Such a suggestion is in 
keeping with the spirit of of that have arisen in the past few years. 
The most famous one is probably SETI@home~\cite{seti}.  SETI@home's 
screensaver downloads and analyzes radio-telescope data.  Others 
similar projects are Folding@home~\cite{folding}, a computation 
application that studies protein folding, and, 
distributed.net~\cite{distributed} a general-purpose distributed 
computing project.  The first publicly downloadable distributed route 
tracing tool is \textsc{\dfn{DIMES}}~\cite{dimes}, released as a 
daemon in September 2004. At the time of writing this report, 
\textsc{DIMES} counts more than 6,000 agents scattered over five 
continents.

However, building such a large structure leads to potential scaling 
issues: the quantity of probes launched might consume undue network 
resources and the probes sent from many vantage points might appear 
as a distributed denial-of-service (DDoS) attack to end-hosts.  These 
problems were quantified in our prior 
work~\cite{DonnetSigmetrics2005}. There are two ways to avoid these
problems: the first one is to stay small, as skitter does for 
instance, but this solution is opposed to the basic idea of scaling 
up the number of tracing monitors. The second is to trace slowly, 
as does \textsc{Dimes}.  In this case, the problem is that the 
resulting network snapshot may be blurred bye the routing changes that
take place over the course of a probing interval.

The Doubletree algorithm~\cite{DonnetSigmetrics2005} is a first 
attempt to perform large-scale topology discovery efficiently and in 
a network friendly manner. Doubletree acts to avoid retracing the 
same routes in the internet by taking advantage of the tree-like 
structure of routes fanning out from a source or converging on a 
destination.  The key to Doubletree is that monitors share 
information regarding the paths that they have explored. If one 
monitor has already probed a given path to a destination then another 
monitor should avoid that path. Probing in this manner can 
significantly reduce load on routers and destinations while 
maintaining high node and link coverage~\cite{DonnetSigmetrics2005}.  
By avoiding redundancy, not only is Doubletree able to reduce the 
load on the network but it also allows one to probe the network more 
frequently. This makes it possible to better capture network 
dynamicity (routing changes, load balancing) compared to standard
approaches based on traceroute.

This report goes beyond earlier theory and simulation to propose a 
Doubletree implementation written in Java~\cite{java}.  We call this 
prototype \dfn{traceroute@home}.  traceroute@home is certainly not 
the first tracerouting tool developed.  However, it differs from 
standard approaches, such as skitter, due to its distributed aspect 
and its scaling resistance thanks to Doubletree.   traceroute@home is 
easily tunable and extensible for further developments.  
traceroute@home is completely open-source and freely 
downloadable~\cite{prototype}.

To validate traceroute@home, we deployed it on the 
PlanetLab~\cite{planetLab} testbed.  We installed traceroute@home on
ten PlanetLab nodes and probed the network, using 200 other PlanetLab
nodes as destinations.

In this report, we describe the traceroute@home system.  We evaluate 
the performance of our prototype, point out its weaknesses and the 
problems encountered in its deployment. We discuss directions for 
future development of our tool and the opportunity for creating an 
infrastructure entirely dedicated to network measurement.

The remainder of this report is organized as follows: 
Sec.~\ref{traceroute} explains how traceroute works; Sec.~\ref{dt} 
describes the Doubletree algorithm and some of its extensions; 
Sec.~\ref{impl} describes traceroute@home; in Sec.~\ref{deployment}, 
we discuss its on PlanetLab nodes; Sec.~\ref{further} introduces 
directions for future extensions; finally, Sec.~\ref{conclusion} 
summarizes the principal contributions of this report.

\section{Traceroute}\label{traceroute}
%%%%%%%%%%%%%%%%%%%%%
Traceroute is a networking tool that allows one to discover the path 
a data packet takes to go from a machine S (the \dfn{source} or the 
\dfn{monitor}) to a machine D (the \dfn{destination}). 

\begin{figure}[!t]
  \begin{center}
    \includegraphics[width=6cm]{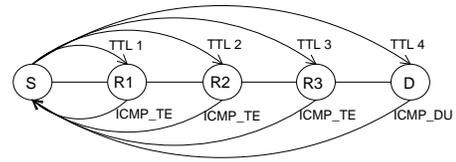}
  \end{center}
  \caption{Traceroute example}
  \label{traceroute.example}
\end{figure}

Fig.~\ref{traceroute.example} illustrates how traceroute works. 
\texttt{S} is the source of the traceroute, \texttt{D} is the 
destination and the \texttt{Ri}s are the routers along the path. S 
sends multiple UDP probes, UDP being the \dfn{User Datagram Protocol} 
which is a connectionless transport protocol, into the network with 
increasing \dfn{time-to-live} (TTL) values, the TTL being a field in 
the IP header indicating how long a packet can circulate in the 
network. Each time a packet enters a router, the router decrements 
the TTL.  When the TTL value is one, the router determines that the 
packet has consumed sufficient resources in the network, drops it, 
and informs the source of the packet by sending back an ICMP 
\textit{time exceeded} message (\texttt{ICMP\_TE} in 
Fig.~\ref{traceroute.example}).  ICMP, Internet Control Message 
Protocol, is a protocol for managing errors related to networked 
machines.  By looking at the IP source address of the ICMP message, 
the monitor can learn the IP address of the router at which the probe 
packet stopped.

When, finally, a probe reaches the destination, the destination is 
supposed to reply with an ICMP \textit{destination unreachable} 
message (\texttt{ICMP\_DU} in Fig.~\ref{traceroute.example}).

Unfortunately, the traceroute behavior explained above is the ideal 
case.  A router along the path might not reply to probes because the 
ICMP protocol is not activated, or the router is overloaded. In order 
to avoid waiting an infinite time for the ICMP reply, the traceroute 
monitor actives a timer when it launches the UDP probe.  If the timer 
expires and no reply was received, then, for that TTL, the machine is 
considered as \textit{non-responding}. 

However, a particular problem occurs when it is the destination that 
does not reply to probes because, for instance, of a restrictive 
firewall.  In this case, the destination will be recorded as non 
responding but it is impossible to know that it was reached.  In 
order to avoid inferring a boundless path, an upper bound on the 
number of successive non-responding machines is used.  For instance, 
in skitter and in our application, this upper bound is set to five.

Standard traceroute, as just described, is based on UDP probes. 
However, two variants exist.  The behavior of the traceroute for the 
intermediate routers is the same as standard traceroute.  The 
difference comes with the destination. The first variant is based on 
ICMP.  Instead of launching UDP probes, the source sends ICMP 
\textit{Echo Request} messages.  The destination is supposed to reply 
with an ICMP \textit{Echo Reply}.  The second sends packets using the 
\dfn{Transport Control Protocol} (TCP) which is a connection-oriented 
transport protocol.  The TCP traceroute aims to bypass most common 
firewall filters by sending TCP SYN packets.  It assumes that 
firewalls will permit inbound TCP packets to specific ports listening 
for incoming connections.

\section{Doubletree}\label{dt}
\begin{figure}[!t]
  \begin{center}
    \subfigure[Monitor-rooted]{\label{dt.tree.monitor}
      \includegraphics[width=5.5cm]{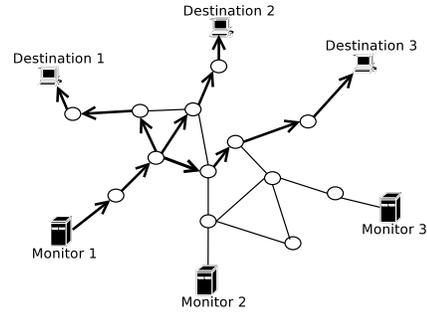}}
    \subfigure[Destination-rooted]{\label{dt.tree.destination}
      \includegraphics[width=5.5cm]{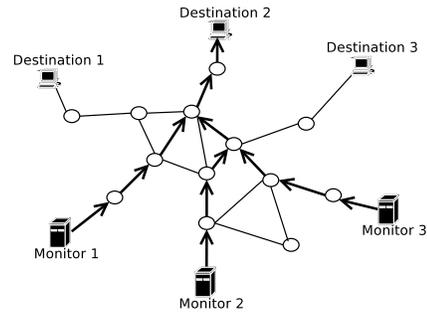}}
  \end{center}
  \label{dt.tree}
  \caption{Tree-like routing structures}
\end{figure}

%%%%%%%%%%%%%%%%%%%%
Doubletree~\cite{DonnetSigmetrics2005} is the key component of a 
coordinated probing system that significantly reduces load on routers 
and end-hosts while discovering nearly the same set of nodes and 
links as standard approaches based on traceroute. It takes advantage 
of the tree-like structures of routes in the context of probing. 
Routes leading out from a monitor towards multiple destinations form 
a tree-like structure rooted at the monitor (see 
Fig.~\ref{dt.tree.monitor}). Similarly, routes converging towards a 
destination from multiple monitors form a tree-like structure, but 
rooted at the destination (see Fig.~\ref{dt.tree.destination}). A 
monitor probes hop by hop so long as it encounters previously unknown 
interfaces. However, once it encounters a known interface, it stops, 
assuming that it has touched a tree and the rest of the path to the 
root is also known.  Using these trees suggests two different probing 
schemes: backwards (monitor-rooted tree) and forwards 
(destination-rooted tree).

For both backwards and forwards probing, Doubletree uses stop sets.  
The one for backwards probing, called the \dfn{local stop set}, 
consists of all interfaces already seen by that monitor.  Forwards 
probing uses the \dfn{global stop set} of $(\mathrm{interface}, 
\mathrm{destination})$ pairs accumulated from all monitors.  A pair 
enters the stop set if a monitor received a packet from the interface
in reply to a probe sent towards the destination address.

A monitor that implements Doubletree starts probing for a destination 
at some number of hops $h$ from itself. It will probe forwards at 
$h+1$, $h+2$, etc., adding to the global stop set at each hop, until 
it encounters either the destination or a member of the global stop 
set.  It will then probe backwards at $h-1$, $h-2$, etc., adding to 
both the local and global stop sets at each hop, until it either has 
reached a distance of one hop or it encounters a member of the local 
stop set. It then proceeds to probe for the next destination.  When 
it has completed probing for all destinations, the global stop set is 
communicated to the next monitor.

Doubletree has one tunable parameter. The choice of initial probing 
distance $h$ is crucial.  Too close, and duplication of effort will 
approach the high levels seen by classic forwards probing 
techniques~\cite[Sec.~2]{DonnetSigmetrics2005}. Too far, and there 
will be high risk of traffic looking like a DDoS attack for 
destinations.  The choice must be guided primarily by this latter 
consideration to avoid having probing look like a DDoS attack.

\begin{figure}[!t]
  \begin{center}
    \includegraphics[width=6.5cm]{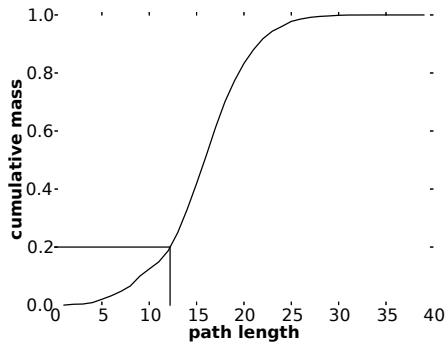}
  \end{center}
  \label{dt.path}
  \caption{Cumulative mass plot of path lengths from skitter monitor
    \texttt{apan-jp}}
\end{figure}

While Doubletree largely limits redundancy on destinations once 
hop-by-hop probing is underway, its global stop set cannot prevent 
the initial probe from reaching a destination if $h$ is set too high. 
Therefore, each monitor sets its own value for $h$ in terms of the 
probability $p$ that a probe sent $h$ hops towards a randomly 
selected destination will actually hit that destination. 
Fig.~\ref{dt.path} shows the cumulative mass function for this 
probability for skitter monitor \texttt{apan-jp}.   If one considers 
as reasonable a $0.2$ probability of hitting a responding destination 
on the first probe, it must chose $h \leq 12$.

Simulation results~\cite[Sec.~3.2]{DonnetSigmetrics2005} show for a 
range of $p$ values that, compared to classic probing, Doubletree is 
able to reduce measurement load by approximately 70\% while 
maintaining interface and link coverage above 90\%.

However, one possible obstacle to successful deployment of Doubletree 
concerns the communication overhead from sharing the global stop set 
among monitors. Tracing from 24 monitors, i.e., the same quantity of 
probing monitors as skitter, to a relatively small set of just 50,000 
destinations with $p=0.05$ produces a set of 2.7 million 
$(\mathrm{interface}, \mathrm{destination})$ pairs.  As pairs of IPv4 
addresses are 64 bits long, an uncompressed stop set based on these 
parameters requires 20.6 MB.

A way to reduce this communication overhead is to use \dfn{Bloom 
filters}~\cite{DonnetPam2005} to implement the global stop set.  A 
Bloom filter~\cite{bloom} summarizes information concerning a set in 
a bit vector that can then be tested for set membership.  An empty 
Bloom filter is a vector of all zeroes. A key is registered in the 
filter by hashing it to a position in the vector and setting the bit 
at that position to one. Multiple hash functions may be used, setting 
several bits to one. Membership of a key in the filter is tested by 
checking if all hash positions are set to one. A Bloom filter will 
never falsely return a negative result for set membership. It might, 
however, return a false positive. For a given number of keys, the 
larger the Bloom filter, the less likely is a false positive. The 
number of hash functions also plays a role.

In prior work~\cite{DonnetPam2005}, we have shown that, when 
$p=0.05$, using a bit vector of size $10^7$ and five hash functions 
allows nearly the same coverage level as a list implementation of the 
global stop set while slightly reducing the redundancy on both 
destinations and internal interfaces and yielding a compression 
factor of 17.3.

Donnet and Friedman, co-authors of this report, also proposed an 
enhancement to the forwards stopping rule based on Classless 
Inter-Domain Routing (CIDR) address prefixes~\cite{DonnetEunice2005}. 
The idea is to aggregate the destinations set by recording the CIDR 
address prefixes of destinations rather than the full IP address. 
This allows one to reduce the amount of communication required by 
Doubletree.  Instead of sharing a set of $(\mathrm{interface}, 
\mathrm{destination})$ pairs, monitors will share a set of 
$(\mathrm{interface}, \mathrm{prefix\_destination})$ pairs.  The 
shortest the prefix, the more destinations can be represented by a 
single entry, but also the more likely the entry will generate false 
positives.  With this simple mechanism, load on destinations can be 
further reduced while maintaining the coverage accuracy around 90\%. 
When combined with a Bloom filter, one further reduces the global 
stop set size, providing a compression factor of 57.1.

\section{traceroute@home Implementation}\label{impl}
%%%%%%%%%%%%%%%%%%%%%%%%%%%%%%%%%%%%%%%%%
This section describes the traceroute@home implementation. 
Sec.~\ref{impl.design} presents our design choices. 
Sec.~\ref{impl.global} introduces the macroscopic functioning of the 
cooperative system that makes use of the Doubletree algorithm. 
Sec.~\ref{impl.monitor} takes a microscopic look at traceroute@home 
by explaining the behavior of a monitor and each module composing it. 
Sec.~\ref{impl.msg} discusses the general messages framework.

\subsection{Design Choices}\label{impl.design}
%%%%%%%%%%%%%%%%%%%%%%%%%%%%
We implemented traceroute@home in Java~\cite{java}. We choose Java as 
the development language because of two reasons: the large quantity 
of available packages and the possibility of abstracting ourselves 
from technical details.  As a consequence, the development time was 
strongly reduced.  Unfortunately, Sun does not provide any package 
for accessing packet headers and handling raw sockets, which is 
necessary to implement traceroute. Instead of developing our own raw 
sockets library, we used the open-source \dfn{JSocket Wrench} 
library~\cite{jsocket}.  We modified the JSocket Wrench library in 
order to support multi-threading.  Our modifications are freely
available~\cite{prototype}. 

We aimed for the design of traceroute@home to be easily extended in 
the future by ourselves but also by the networking community.  For 
instance, concerning the messages exchanged by monitors, we define a 
general framework for messages, making creation and handling of new 
messages easier.  In addition to that, traceroute@home is readily 
tunable due to a configuration file that is loaded by the application 
at its starting.  Our implementation is freely 
available~\cite{prototype}.

We designed our application by considering two levels: the 
\dfn{microscopic} level and the \dfn{macroscopic} level.

From a macroscopic point of view, i.e., all the monitors together, 
the monitors are organized in a ring, adopting a round robin process. 
At a given time, each monitor focuses on its own part of the 
destination list.  When it finishes probing its part, it sends 
information to the next monitor and waits for data from the previous 
one, if it was not yet received.  Sec.~\ref{impl.global} explains 
this macroscopic aspect of traceroute@home.

From a microscopic point of view of our implementation, i.e., a single
monitor, a monitor is composed of several modules that
interact with each other.  Our implementation is thread-safe, as a
monitor is able to send several probes at the same time.  Further,
topological information collected by a monitor is regularly saved to
XML files.  Sec.~\ref{impl.monitor} explains this microscopic level of
traceroute@home. 

\subsection{System Overview}\label{impl.global}
%%%%%%%%%%%%%%%%%%%%%%%%%%%%%
The simulations conducted in prior work~\cite{DonnetSigmetrics2005} 
were based on a simple probing system: each monitor in turn covers 
the destination list, adds to the global stop set the 
$(\mathrm{interface}, \mathrm{destination})$ pairs that it 
encounters, and passes the set to the subsequent monitor.

This simple scenario is not suitable in practice: it is too slow, as 
an iterative approach allows only one monitor to probe the network at 
a given time.  We want all the monitors probing in parallel.  
However, how would one manage the global stop set if it were being 
updated by all the monitors at the same time?

\begin{figure}[!t]
  \begin{center}
    \includegraphics[width=8cm]{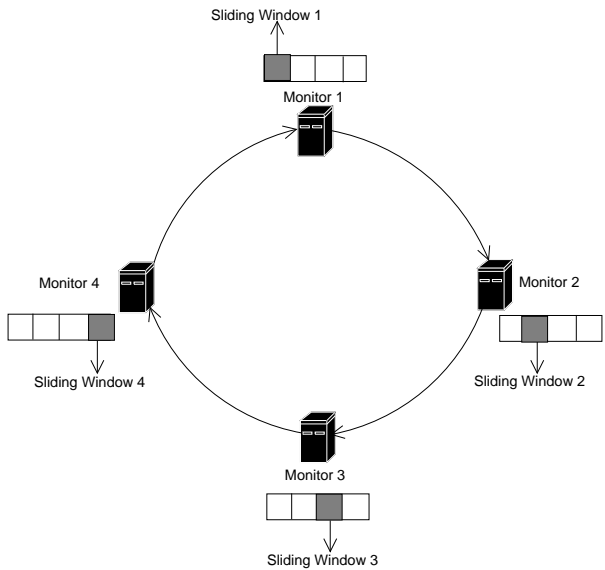}
  \end{center}
  \caption{Doubletree with sliding windows}
  \label{impl.global.window}
\end{figure}

An easy way to parallelize is to deploy several \dfn{sliding 
windows} that slide along the different portions of the destination 
list. At a given time, a given monitor focuses on its own window, as 
shown in Fig.~\ref{impl.global.window}. There is no collision between 
monitors, in the sense that each one is filling in its own part of 
the global stop set.  The entire system counts $m$ different sliding 
windows, where $m$ is the number of Doubletree monitors.  If there 
are $n$ destinations, each window is of size $w = n/m$.  This is an
upper-bound on the window size as the concept still applies if they
are smaller.

A sliding window mechanism requires us to decide on a step size by 
which to advance the window.  We could use a step size of a single 
destination.  After probing that destination, a Doubletree monitor 
sends a small set of pairs corresponding to that destination to the 
next monitor, as its contribution to the global stop set.  It 
advances its window past this destination, and proceeds to the next 
destination.  Clearly, though, a step size of one will be costly in 
terms of communication. Packet headers (see Sec.~\ref{impl.msg} for 
details about packet format) will not be amortized over a large 
payload, and the payload itself, consisting of a small set, will not 
be as susceptible to compression as a larger set would be.

On the other hand, a step size equal to the size of the window itself
poses other risks.  Suppose a monitor has completed probing each
destination in its window, and has sent the resulting subset of the
global stop set on to the following monitor.  It then might be in a
situation where it must wait for the prior monitor to terminate its
window before it can do any further useful work.

A compromise must be reached, between lowering communications costs 
and continuously supplying each monitor with useful work.  This 
implies a step size somewhere between 1 and $w$.  For our 
implementation of Doubletree, we let the user decide the step size. 
This is a part of the XML configuration file that each Doubletree 
monitor loads at its start-up (see Sec.~\ref{impl.monitor}).   Future 
work might reveal information about how to tune the step size of a 
monitor.

\subsection{Inside a traceroute@home Monitor}\label{impl.monitor}
\begin{figure*}[!t]
  \begin{center}
    \includegraphics[width=10cm]{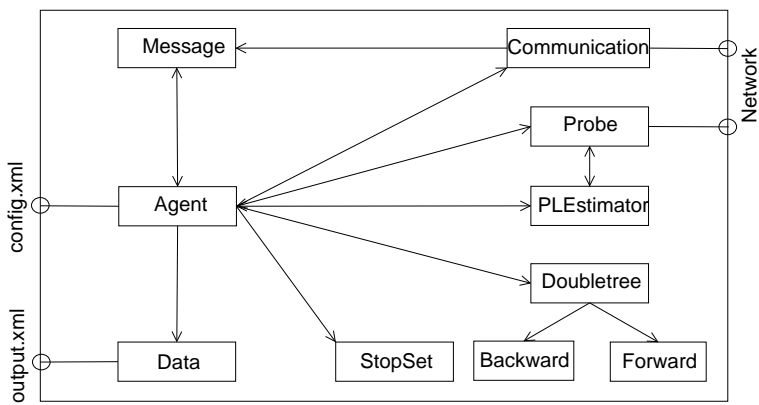}
  \end{center}
  \caption{A traceroute@home monitor's modules}
  \label{impl.monitor.fig}
\end{figure*}

%%%%%%%%%%%%%%%%%%%%%%%%%%%%%%%%%%%%%%%%%%%%%%
Fig.~\ref{impl.monitor.fig} shows the different modules composing a
traceroute@home monitor and the way they interact with each other (the
arrows) and with their environment, i.e., hard disk or network (circles).

First, a traceroute@home monitor loads an XML file 
(\textit{config.xml}) that contains configuration information, such 
as the number of sliding windows and the name (or IP address) of the 
next monitor in the round robin process.  This XML file must strictly 
follow a DTD (Document Type Definition).\footnote{For interested 
readers, the DTD is available online~\cite{prototype}.} In the 
current version of the application, the XML file must be present on 
the machine running the software. In the following versions (See 
Sec.~\ref{further}), we can imagine that a traceroute@home monitor 
will upload this file from a remote server. Note also that the
destination list and sliding window information must also be present 
on the machine.  We can envisage that these will also be remotely
available in the future.

This configuration file is processed by the \dfn{Agent}.  The Agent 
is the heart of a traceroute@home monitor as it first creates all 
other modules, manages them and, finally, allows the various 
components to interact with each other.

The Agent creates the \dfn{StopSet} module that implements the stop 
set data structure~\cite{DonnetSigmetrics2005}.  Our stop set 
implementation is multithread safe. Two types of implementations are 
proposed: list and Bloom filter.  The list is the basic 
implementation and can be used for the local stop set as well as the 
global stop set.  The Bloom filter 
implementation~\cite{DonnetPam2005} may only be used for the global 
stop set. The hash functions needed by the Bloom filter are emulated 
with the SHA-1 algorithm~\cite{sha1}. Depending on the XML 
configuration file, the global stop set may be compressed or not 
before being sent to the next monitor.  Note that, for consistency 
reasons, each monitor in the system must use the same implementation 
for the global stop set.

Doubletree is a cooperative algorithm.  The different monitors have 
to share their global stop set.  They thus exchange messages.  The 
purpose of the \dfn{Message} component is therefore to build, parse 
and store messages (before their handling by the Agent) sent and 
received by a monitor.  We explain in detail the Message component 
by describing our general message framework in Sec.~\ref{impl.msg}.

A traceroute@home monitor, through the Message module, is able to 
handle messages.  To send and receive them, it makes use of the 
\dfn{Communication} component that allows a monitor to interact with 
other monitors.  It sends messages to a given monitor when the Agent 
orders it and listens to potential connections for incoming messages. 
Incoming messages are parsed and stored by the Message module before 
their handling by the Agent. In order to make this module as 
efficient as possible, it was implemented using a \dfn{selector}. 
A selector provides the ability to do readiness selection, which 
enables multiplexed I/O operations. A selector makes it possible for a
 single thread to manage many I/O channels simultaneously.  This 
corresponds to the \texttt{select()} operation in C.

The \dfn{Probe} module builds probes (UDP or ICMP), assigns them a TTL
value, sends them into the network and waits for eventual ICMP replies. 
The Probe module is created by a client that wants to probe the 
network. An example of such a client is the Agent.  The client may 
require multithreading in its network exploration.  Therefore, the 
Probe module allows one to send multiple probes at the same time. 
Another unique thread listens to incoming ICMP replies messages and 
dispatches them to the client. The client is in charge of the 
matching between a probe sent and the eventual ICMP reply.  This 
matching is possible because the ICMP reply (\texttt{destination 
unreachable} or \texttt{time exceeded}) contains the IP header and 
the 8 first bytes of the original datagram (refer to 
Sec.~\ref{traceroute} for details about how traceroute works). 
Placing a unique source port number in the UDP header of each 
outgoing probe allows the returning ICMP replies to be identified. 
Moors discusses the reasons for varying the source port of UDP 
datagram instead of the destination 
port~\cite[Sec.~III]{streamlining}.

The \dfn{PLEstimator} (for ``Path Length Estimator'') module is in 
charge of discovering path lengths for the current sliding window.  
It sends to destinations UDP probes with a TTL of 64 and a high 
destination port.  If a destination replies with a 
\texttt{Destination Unreachable} ICMP message, the PLEstimator is 
able to know the distance by looking at the TTL field contained in 
the IP header copied in the ICMP message payload.  A simple 
subtraction allows one to know the path length. With all the path 
length information received, the PLEstimator builds the path length 
CDF, as shown in Fig.~\ref{dt.path}.  As suggested in prior 
work~\cite{DonnetSigmetrics2005}, the PLEstimator will use, by 
default, a $p$ value of 0.05 in order to determine the $h$ value that 
must be used for the current sliding window.  The PLEstimator is a 
client of the Probe module.

The \dfn{Doubletree} module defines an interface describing the 
general behavior of a probing scheme.  This interface is implemented 
in two ways, such that it defines the two probing scheme behaviors, 
i.e., backwards and forwards, proposed by prior 
work~\cite{DonnetSigmetrics2005}. However, the Doubletree module is 
not a Probe module client. In this case, multithreading is managed by 
the Agent.  The Doubletree module is only used to decide, according 
to a probe reply (which might be empty if the router did not reply to 
probes), if a stopping condition is reached.  Four stopping 
conditions are defined: \textit{i)} normal (the destination, forwards 
probing, or the first hop, backwards probing, is reached), 
\textit{ii)} stop set, \textit{iii)} loop and \textit{iv)} gap.  A 
gap occurs when the monitor encounters five successive non-responding 
interfaces. The Doubletree module is also in charge of deciding the 
next TTL value.

When a traceroute@home monitor has finished probing a part of the 
sliding window corresponding to the step size, it sends the 
corresponding stop set to the subsequent monitor but it also saves 
the topological information that it has gathered.  Maintaining 
information about the topology discovered and saving this information 
on the hard disk is the \dfn{Data} module's job.  The topological 
information is transformed into an XML file according to a 
DTD.\footnote{Interested readers may find it online~\cite{prototype}.} 
The XML format was chosen in order to facilitate eventual later 
data migration into a more complex data type.  An XML format also 
makes the data handling easier.  For each destination probed, we 
currently record the following information: the source (i.e., the IP 
address of the monitor), the destination, a timestamp, the backwards 
probing stopping reason as described above, the backwards probing 
stopping distance, the forwards probing stopping reason, the forwards 
probing stopping distance, and the path discovered.  The path is 
composed of several hops, a hop being composed of the TTL value and 
three IP addresses or ``*'', corresponding to a non-responding 
interface. If the IP address corresponds to a responding router, then 
the round-trip time (RTT) for the probe and the response is added.

\subsection{Message Exchange}\label{impl.msg}
%%%%%%%%%%%%%%%%%%%%%%%%%%%%%%
Monitors in the system have to share information about what was 
previously discovered.  This is achieved by regularly exchanging 
their global stop set.

A trivial implementation could have a monitor simply send its global 
stop set as a byte stream to another monitor, without any additional 
information. However, such a simple mechanism would make further 
extensions difficult, specially in the case of a peer-to-peer or 
overlay system used to manage the whole system (see 
Sec.~\ref{further}).

For that reason, we provide a general framework for messages 
exchanged between monitors.  This framework might be easily extended 
in order to manage any type of messages.

In this section, we describe the message framework we propose and a 
specific message that is actually the only one currently implemented: 
\textsc{StopSet}. Both will be first described in Augmented 
Backus-Naur Form (ABNF)~\cite{abnf} and second as a byte stream.

A message in a traceroute@home system is composed of the two parts: 
the \dfn{header} and the \dfn{payload}.  The header is mandatory and 
gives information about the message length and its type.  As opposed 
to the header, the payload is optional.  Some messages, for instance 
a heartbeat-like message, do not need any payload.  The ABNF for a 
message is given in Fig.~\ref{impl.msg.msgabnf}.

\begin{figure}[!t]
  \begin{center}
    \begin{grammar}
      [(colon){$=$}] 
  	  [(semicolon)$|$] 
  	  <msg> : <header><payload>$^?$\\ 
  	  <header> : <length><type>\\ 
  	  <length> : <integer$_{16}$>\\ 
  	  <type> : <stopset>\\ 
  	  <stopset> : "0"\\ 
  	  <integer$_{16}$> : "4" ; \ldots ; "65535"
	\end{grammar}
  \end{center}
  \caption{ABNF for a message}
  \label{impl.msg.msgabnf}
\end{figure}

\begin{figure}[!t]
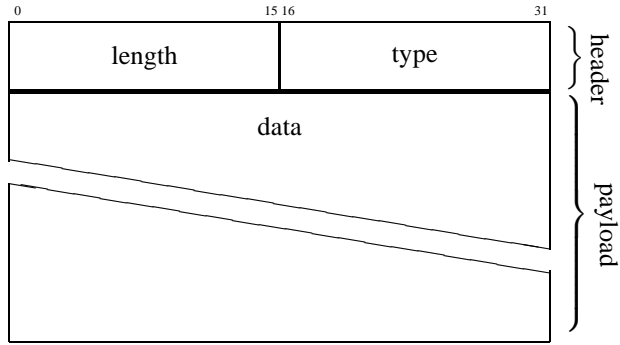

  \begin{flushright}
  \setlength{\byteheight}{6ex}
  \begin{bytefield}{32}
    \bitheader{0, 15-16, 31}\\
    \wordgroupr{\rotatebox{-90}{header}}
      \bitbox{16}{length} & \bitbox{16}{type}
    \endwordgroupr\\    
	\wordgroupr{\rotatebox{-90}{payload}}
	  \wordbox[lrt]{1}{data}\\
	  \skippedwords\\
	  \wordbox[lrb]{1}{}
	\endwordgroupr\\
  \end{bytefield}
  \end{flushright}
  \caption{General packet format}
  \label{impl.msg.packet}
\end{figure}

Fig.~\ref{impl.msg.packet} provides a byte-stream vision of a 
message. Length and type are expressed as 16 bit integers. Note that
the length includes both the header and the payload.  The minimum
length of a message is thus four bytes, i.e., the header length.  The
type and length encoding is big-endian.

In the current traceroute@home implementation, a monitor sends and 
receives only one type of message: the \textsc{StopSet}. This message 
contains information about the topology discovered by a monitor.  A 
\textsc{StopSet} message is sent when, for a given sliding window, a 
step size is reached.  Two pieces of information (number of the 
sliding window and the step size considered) must be present in the 
\textsc{StopSet} message so that the receiver might identify the 
message and link it to a portion of a sliding window. In the ABNF, 
the step size is called a \dfn{slice}. A \textsc{StopSet} message 
must provide information on implementation details: the type of stop 
set (the \dfn{stype}, list or Bloom filter), the type of network 
probed (the \dfn{ip}, IPv4 or IPv6)\footnote{Currently, only IPv4 is 
implemented. } and the eventual compression (the \dfn{compress}) of 
the stop set.  Of course, it also contains the relevant portion of 
the global stop set. The \textsc{StopSet} payload in an ABNF format 
is given by Fig.~\ref{impl.msg.ssabnf}.

\begin{figure}[!t]
  \begin{center}
	\begin{grammar}
  	  [(colon){$=$}]
  	  [(semicolon)$|$]
  	  <payload> : <window><slice><impl>\\<stopset>\\
  	  <window> : <integer$_8$>\\
  	  <slice> : <integer$_8$>\\
  	  <impl> : <stype><ip><compress>\\
  	  <stype> : <bit>\\
  	  <ip> : <bit>\\
  	  <compress> : <bit>\\
  	  <stopset> : <byte>$^+$\\
  	  <integer$_8$> : "1" ; \ldots ; "32767"\\
  	  <byte> : <bit>*8\\
  	  <bit> : "0" ; "1"
    \end{grammar}
  \end{center}
  \caption{ABNF for the \textsc{StopSet} message}
  \label{impl.msg.ssabnf}
\end{figure}

\begin{figure}[!t]
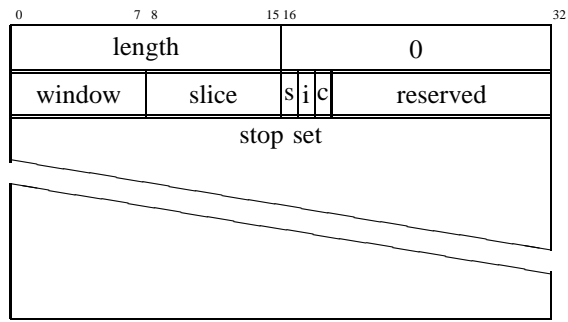

  \begin{center}
    \begin{bytefield}{32}
	  \bitheader{0, 7-8, 15-16, 32}\\
	  \bitbox{16}{length} & \bitbox{16}{0}\\
	  \bitbox{8}{window} & \bitbox{8}{slice} & \bitbox{1}{s} & \bitbox{1}{i} & \bitbox{1}{c} & \bitbox{13}{reserved}\\
	  \wordbox[lrt]{1}{stop set}\\
	  \skippedwords\\
	  \wordbox[lrb]{1}{}
	\end{bytefield}
  \end{center}
  \caption{\textsc{StopSet} packet format}
  \label{impl.msg.stopset}
\end{figure}

Fig.~\ref{impl.msg.stopset} shows the \textsc{StopSet} message as a
byte stream.  We see that 13 bits are reserved for future extensions. 
Padding is used to fill in this unused part of the packet.  The stop
set is encoded as a byte array.  If the stop set is implemented as a
list, a group of 8 bytes refers to a global stop set key.  The first
four bytes represent the interface address and the last four bytes
represent the destination address.  The principle is identical modulo
group size in case of IPv6 addresses.

\section{Deployment Story}\label{deployment}
%%%%%%%%%%%%%%%%%%%%%%%%%%
This section talks about the deployment and validation of 
traceroute@home on PlanetLab nodes, the different difficulties we 
encountered (not necessarily related to PlanetLab) and the way we 
solved them (Sec.~\ref{deployment.prob}). 
Sec.~\ref{deployment.success} presents our deployment results. 

\subsection{Difficulties Encountered}\label{deployment.prob}
%%%%%%%%%%%%%%%%%%%%%%%%%%%%%%%%%%%%%%%
Some routers along the path may be poorly configured.  It seems that 
they, when building the ICMP message, can modify the original 
datagram. Several ICMP messages were returned with the source and 
destination ports changed in the original datagram.  This is a 
critical issue as the source port of the originating UDP datagram is 
different for each datagram, as explained in Sec.~\ref{impl.monitor}, 
in order to identify the thread that sends the datagram.  This 
problem can be avoided by also checking the destination address in 
the original IP header.  

By definition, a PlanetLab node is minimalist in the sense that it 
provides a nearly empty file system.  The only environments provided 
consist of Perl (version 5.8.3) and Python (version 2.3.3). By 
default, there are no compilation possibilities (no make, no gcc, no 
g++) and no Java environment.  We had to install a Java runtime 
environment, on each nodes supposed to run traceroute@home.  
%An 
%automatic package loading system would be a nice feature for eventual 
%inclusion in PlanetLab.

Table~\ref{deployment.prob.reach} describes the availability of 
PlanetLab nodes in December, 2005. \textit{Offline} nodes are those 
that are either being installed or having long term issues.  When 
gathering these statistics, 18.9\% of the PlanetLab nodes were 
offline. \textit{unreachable} nodes are in production, i.e., the 
PlanetLab system is running, but not reachable via SSH.  10.9\% of 
the nodes were unreachable. \textit{broken} nodes are those that have 
failed tests but that can be logged into via SSH as root.  No 
PlanetLab nodes were broken. \textit{ok} nodes are those that can be 
used normally. 70.2\% of the PlanetLab nodes were up but it was
important for us to check availability before running any experiment. 

\begin{table}[!t]
  \begin{center}
    \begin{tabular}{r|r}
      \textbf{Usability} & \textbf{Number}\\
      \hline
      offline     & 120\\
      unreachable & 69\\
      broken      & 0\\
      ok          & 448\\
      \hline
      \textbf{total} & 637\\
    \end{tabular}
  \end{center}
  \caption{PlanetLab nodes availability, December 2005}
  \label{deployment.prob.reach}
\end{table}

\begin{figure}[!t]
  \begin{center}
    \includegraphics[width=6.5cm]{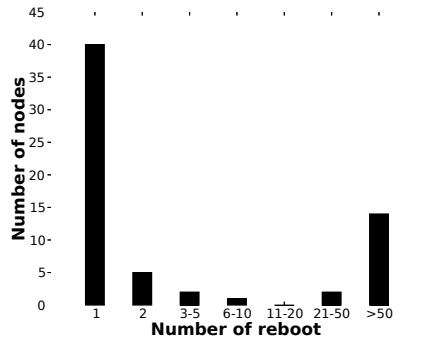}
  \end{center}
  \caption{PlanetLab nodes reboot, one week (December 2005)}
  \label{deployment.prob.reboot}
\end{figure}

PlanetLab nodes reboot periodically for scheduled maintenance and 
upgrades.  Fig.~\ref{deployment.prob.reboot} shows the number of 
reboot during one week in December 2005.  64 PlanetLab nodes were 
involved in rebooting, for a total of 18,577 reboots.  Among these 64 
nodes, 40 rebooted only once.  However, 14 nodes rebooted more than 
50 times.  Among these 14 nodes, 10 nodes rebooted more than 1,000 
times during the one week period and one node rebooted 3,144 times. 
As it is not clear how to know which PlanetLab node will reboot and 
when, a long-term experiment must allow for the possibility that a 
portion of the nodes will reboot.

\begin{figure}[!t]
  \begin{center}
    \includegraphics[width=6.5cm]{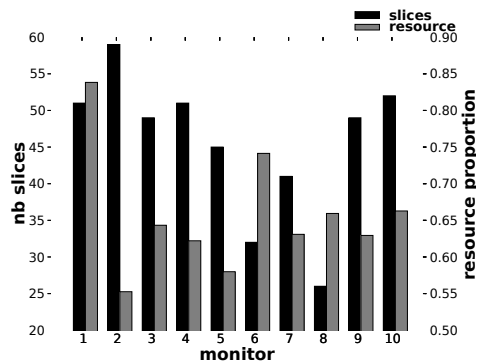}    
  \end{center}
  \caption{traceroute@home monitors evaluation (December \nth{20} 2005)} 
  \label{deployment.prob.bench}
\end{figure}

Fig.~\ref{deployment.prob.bench} evaluates the performances of the 
ten PlanetLab nodes we used as traceroute@home monitors (see 
Sec.~\ref{deployment.success} for details about the traceroute@home 
monitors). The performance statistics were gathered on December 
\nth{20}, 2005.  The horizontal axis shows the PlanetLab nodes we 
used as traceroute@home monitors.  The left-side vertical axis gives 
the number of active slices on each monitor (black bar).  A slice is 
the PlanetLab term for an account. The right-side vertical axis 
gives the proportion of network resources used by the most consuming 
slice on each monitor (grey bar).  By network resources, we understand
the quantity of information sent and received.

Regarding first the quantity of slices per monitor, we see that the
maximum, 59, is reached with the monitor labeled 2.  The minimum is
32 for the monitor 6.  On average, a PlanetLab node chosen for being a
traceroute@home monitor hosted 45 slices.  These statistics give us an
idea of a PlanetLab node load, as each PlanetLab node is supposed to
affect resources to a slice in a best effort way.

The right-hand of the vertical axis informs us of the proportion of
network resources used, on each monitor, by the most active slice.  It
oscillates between 0.5526 (monitor 2) and 0.8383 (monitor 1).  On
average, 0.65601 of the network resources are used by a single slice. 
%It means thus that the remaining 34\% of the resources must be shared
%by the other slices.

\subsection{Deployment Success}\label{deployment.success}
%%%%%%%%%%%%%%%%%%%%%%%%%%%%%%%%
As described in prior work~\cite{DonnetSigmetrics2005}, security
concerns are paramount in large-scale active probing.  It is important
to not trigger alarms inside the network with Doubletree probes.  It
is also important to avoid burdening the network and the destination
hosts. It follows from this that the deployment of a cooperative
active probing tool must be done carefully, proceeding step by step,
from an initial small size, up to larger-scales.  Note that this
behavior is strongly recommended by PlanetLab~\cite[Pg.~5]{plMyth}.

Our application was deployed to only ten PlanetLab nodes.  We selected
ten nodes based on their relatively high stability (i.e., remaining up
and connected), and their relatively low load.  These traceroute@home
monitors are scattered around the world: North America (USA, Canada),
Europe (France, Spain, Switzerland, Spain), and Asia (Japan, Korea). 
In the future, we will wish to scale up the number of monitors to, at
least, the skitter scale (i.e., 24 monitors).  

The destination list consists of $n=200$ PlanetLab nodes randomly 
chosen amongst the approximately 300 institutions that currently host 
PlanetLab nodes.  Restricting ourselves to PlanetLab nodes 
destinations was motivated by security concerns.  By avoiding tracing 
outside the PlanetLab network, we avoid disturbing end-systems that 
do not welcome probe traffic.  None of the ten PlanetLab monitors (or 
other nodes located at the same place) belongs to this destination 
list. The sliding window size of $w=n/m$ consists of twenty 
destinations. We consider two step sizes (i.e., slices) by window, so 
each slice counts ten destinations.

Finally, each traceroute@home monitor was configured as follows: the 
probability $p$ was set to 0.05, the global stop set implementation 
was the list (i.e., the standard implementation) and no compression 
was required before sending the \textsc{StopSet} messages.

The experiment was run on the PlanetLab nodes on Dec. \nth{20} 2005.
All the traceroute@home monitors were started at the same time. The
experiment was finished when each monitor had probed the entire
destination list.

A total of 2,703 links and 2,232 nodes were discovered.  We also
encountered 2,434 non-responding interfaces (routers and
destinations).  We recorded 36 invalid addresses.  Invalid addresses
are, for example, private
addresses~\cite[Sec.~2.1]{DonnetSigmetrics2005}. 

\begin{table*}[!t]
  \begin{center}
    \begin{tabular}{l|llll|ll ll|r}
      & \multicolumn{4}{c}{\textbf{Backwards}} & \multicolumn{4}{c}{\textbf{Forwards}}\\
	  \textbf{monitor} & loop & gap & stop set & normal & loop & gap & stop set & normal & \textbf{h}\\ \hline
	  \texttt{Blast}   & 0   & 0 & 99.5 & 0.5 & 2   & 17   & 50   & 31   &	  7\\
	  \texttt{Cornell} & 0   & 0 & 99   & 1   & 0   & 13.5 & 69.5 & 17   & 7\\
	  \texttt{Ethz}    & 1   & 0 & 98.5 & 0.5 & 2   & 10.5 & 52   & 35.5 & 11\\
	  \texttt{Inria}   & 1.5 & 0 & 97.5 & 1   & 1   & \hspace{0.2cm}4    & 67   & 28   & 15\\
	  \texttt{Kaist}   & 0   & 0 & 99   & 1   & 0.5 & 10.5 & 64.5 & 24.5 & 9\\
	  \texttt{Nbgisp}  & 0.5 & 4 & 95   & 0.5 & 3.5 & 30.5 & 22   & 44   & 7\\
	  \texttt{LiP6}    & 0   & 0 & 99.5 & 0.5 & 1   & \hspace{0.2cm}9.5  & 62.5 & 27   & 11\\
	  \texttt{UCSD}    & 0   & 0 & 99.5 & 0.5 & 0   & 10.5 & 60.5 & 29   & 7\\
	  \texttt{Uoregon} & 0   & 0 & 99.5 & 0.5 & 0   & \hspace{0.2cm}7    & 74.5 & 18.5 & 6\\
	  \texttt{Upc}     & 0.5 & 0 & 99   & 0.5 & 1   & 14   & 57.5 & 27.5 & 15\\
	  \hline
	  \textbf{mean} & 0.35 & 0.4 & 98.6 & 0.65 & 0.11 & 12.7 & 58 & 28.2 & 9\\
    \end{tabular}
  \end{center}
  \caption{Stopping reasons (in \%) and $h$ value per monitor}
  \label{deployment.success.stopping}
\end{table*}

Table~\ref{deployment.success.stopping} shows the different reasons
for stopping backwards and forwards probing for each traceroute@home 
monitor.  It further indicates the $h$ value computed by each 
monitor.  The last row of the table indicates the mean for each 
column.

Looking first at the backwards stopping reasons, we see that the
stop set rule strongly dominates (98.6\% on average).  On average, 
normal stopping (i.e., reaching the first hop) occurs only 0.65\% of
the time.

\begin{figure}[!t]
  \begin{center}
    \includegraphics[width=6.5cm]{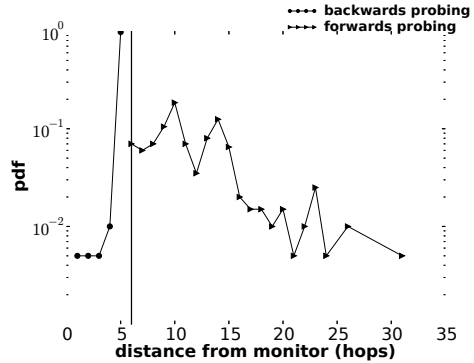}
  \end{center}
  \caption{Stopping distance for the \texttt{Uoregon} monitor}
  \label{deployment.success.distance}
\end{figure}

Fig.~\ref{deployment.success.distance} shows the stopping distance 
(in terms of hops), for a given monitor, \texttt{Uoregon}, when 
probing backwards and forwards. The vertical line indicates the 
$h$ value computed by \texttt{Uoregon}. Results presented in 
Fig.~\ref{deployment.success.distance} are typical for the other
traceroute@home monitors.

We see that more than 90\% of the backwards stopping occurs at a distance
of 5, that is to say the distance corresponding to $h-1$.  In 2.5\% of
the cases, the probing stops between hop 1 (normal stopping) and hop
4.  Except for hop 1, the other stops are caused by the stop set,
probably due to very short paths.  They illustrate the cases in which
the first probe sent with a TTL of $h$ directly hits a destination. 

Looking now at the forwards stopping reasons in
Table~\ref{deployment.success.stopping}, we see that the gap rule
(five successive non responding interfaces) plays a greater rule.  We
believe that these gaps occur when a destination does not respond to
probes because of a restrictive firewall or because the PlanetLab node
is down.  

On average, in 58\% of the cases, the stop set rule applies, and in 
28.2\% of the cases, the normal rule applies.  The normal rule 
proportion might be seen as high but we have to keep in mind 
that a Doubletree monitor starts with an empty stop set.  Therefore, 
during the first sliding window, the only thing that can stop a
monitor, aside from the gap rule, is an encounter with the
destination. 

Looking at the stopping distance in 
Fig.~\ref{deployment.success.distance}, we see that the distances are 
more scattered for forwards probing than for backwards probing. 
Regarding the forwards probing, a peak is reached at a distance of 10 
(18.5\% of the cases).  In 7\% of the cases, the monitor stops 
probing at a distance of 6, that is equal to the value $h$.  It could 
correspond to the stop set rule application or the normal rule, by 
definition of $p$. Recall that $p$ defines the probability of hitting 
a destination with the probe sent with a TTL equals to $h$.  For our 
experiment, we set $p=0.05$, meaning that in 5\% of the cases the 
first probe sent by a monitor will hit a destination.

\begin{table}[!t]
  \begin{center}
    \begin{tabular}{l|r}
      \textbf{monitor} & size\\
      \hline
      \texttt{Blast}   & 12.31\\
      \texttt{Cornell} & 8.77\\
      \texttt{Ethz}    & 7.41\\
      \texttt{Inria}   & 7.85\\
      \texttt{Kaist}   & 11.41\\
      \texttt{Nbgisp}  & 12.84\\
      \texttt{Paris}   & 10.96\\
      \texttt{UCSD}    & 10.44\\
      \texttt{Uoregon} & 11.62\\
      \texttt{Upc}     & 10.29\\
      \hline
      \textbf{mean}    & 10.39\\
    \end{tabular}
  \end{center}
  \caption{Total \textsc{StopSet} message size (in KB) per monitor}
  \label{deployment.success.msg}
\end{table}

Table~\ref{deployment.success.msg} shows the total size of 
\textsc{StopSet} messages (in KB) sent by each monitor.  The size 
takes into account the header of the message (4 bytes) and the 
payload.

A \textsc{StopSet} message is sent by a monitor when it reaches a step
size (i.e. a slice) in the current sliding window.  As we define for
our experiment two step sizes per sliding window and as we deploy our
prototype on ten PlanetLab nodes, each monitor sent 20
\textsc{StopSet} messages.  We tune each Doubletree monitor in order
to use the list implementation of the stopset.

The monitors do not exchange their entire stop set.  They only send an
update that contains the $(\mathrm{interface}, \mathrm{destination})$
pairs discovered during the current step size probing.

In Table~\ref{deployment.success.msg}, we can see that a monitor 
sends a total of between 7.41 KB (\texttt{Ethz}) and 12.84 KB 
(\texttt{Nbgisp}) to the subsequent monitor.  On average, a monitor 
sends 10.39 KB of stop set information into the network.

During our experimentation, the traceroute@home application did not 
flood the network with \textsc{StopSet} messages.  However, our prior 
work~\cite{DonnetPam2005} has shown, on a larger destination list, 
that it can grow to excessive sizes. In this case, we recommend 
configuring a traceroute@home monitor to first use the Bloom filter 
implementation of the stop set and second compress it before sending
it in the network.

\begin{table}[!t]
  \begin{center} 
    \begin{tabular}{l|l|ll}
             		   & \multicolumn{3}{c}{\textbf{time}}\\
      \textbf{monitor} & \textbf{total} & \textbf{probing} & \multicolumn{1}{c}{\textbf{waiting}}\\
      \hline
      \texttt{Blast}   & 24 & 23   & \hspace{0.2cm}1\\
      \texttt{Cornell} & 28 & 15   & 13\\
      \texttt{Ethz}    & 20 & 12.5 & \hspace{0.2cm}7.5\\
      \texttt{Inria}   & 32 & 12.5 & 19.5\\
      \texttt{Kaist}   & 23 & 13   & 10\\
      \texttt{Nbgisp}  & 26 & 10   & 16\\
      \texttt{LiP6}    & 21 & \hspace{0.2cm}9    & 12\\
      \texttt{UCSD}    & 22 & 13.5 & \hspace{0.2cm}8.5\\
      \texttt{Uoregon} & 31 & 31   & \hspace{0.2cm}0\\
      \texttt{Upc}     & 27 & 18.5 & \hspace{0.2cm}8.5\\
	\end{tabular}
  \end{center}
  \caption{Running time (in minutes)}
  \label{deployment.success.time}
\end{table}

\begin{figure}[!t]
  \begin{flushleft}
    \includegraphics[width=8cm]{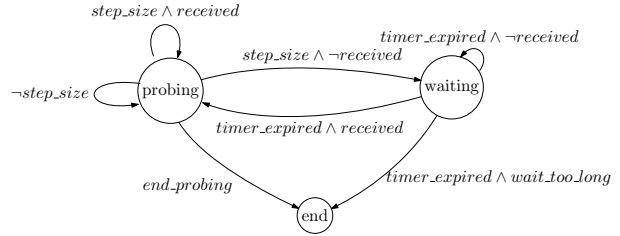}
  \end{flushleft}
  \caption{Probing/waiting state interactions}
  \label{deployment.success.fsm}
\end{figure}

Table~\ref{deployment.success.time} shows, for each traceroute@home 
monitor, the running time (in minutes) in terms of probing and 
waiting.  The waiting period occurs when a monitor has finished its 
sliding window or a slice in a given sliding window and is waiting 
for the global stop set that should be sent by the previous monitor 
in the round-robin topology. We see that nearly all monitors have to 
wait. A waiting period, in our implementation, lasts 30 seconds.  
When the timer expires, the monitor checks if it received a new 
message.  If so, the waiting period ends and a new probing period 
begins. Otherwise, it sleeps during 30 seconds.  To avoid infinite 
waiting, if after 40 sleeping periods (i.e., 20 minutes), nothing was 
received, the monitor quits with an error message. 
Fig.~\ref{deployment.success.fsm} illustrates the interactions 
between the probing state and the waiting state.

We believe that these long waiting periods are due to a 
characteristic of the PlanetLab IP stack.  It seems that when ICMP 
replies are received by the stack, the \texttt{recvfrom}() function 
does not read them immediately. As the timer set on the listening 
socket never expires in this case, we think that the 
\texttt{recvfrom}() function is waiting for the permission to access 
the IP stack.  It looks like the resource is owned (or locked) by 
another process on the PlanetLab node.  Note that this behavior was 
also noticed by other Planet-Lab users~\cite{latency}.

\section{Further Work}\label{further}
%%%%%%%%%%%%%%%%%%%%%%%
This section discusses possible extensions and ways to improve 
traceroute@home (Sec.~\ref{further.dt}). It also discusses some key 
points to enhance measurement infrastructures in general 
(Sec.~\ref{further.infra}).

\subsection{traceroute@home}\label{further.dt}
%%%%%%%%%%%%%%%%%%%%%%%%%%%%%
One of the main aspects we would like to address in the near future 
is the stability of the whole system.  Currently, it is like dominos: 
when a monitor fails, the whole system fails.

From a long term point of view, we aim to develop a peer-to-peer 
(p2p) or overlay application to manage the whole system.  However, it 
is not yet obvious how this p2p/overlay should work.  We need a 
transitional solution.  Currently, our main concern is monitor 
failure recovery.

A simple solution would be to build a centralized server.  We keep 
the basic round robin functioning of the system (see 
Sec.~\ref{impl.global}) but in the middle, we place a server.  In 
this case, the system has a star topology.  The server's job will be to 
maintain the coherency of the round robin structure.

The basic idea is the following: the server knows the entire topology 
of the system and,  for a given monitor, to which monitor it is 
supposed to send its global stop set and from which it must receive 
it.  Regularly, the server checks each monitor's state by sending 
messages of type \textsc{Heartbeat}.  When a monitor receives a 
\textsc{Heartbeat} message, it is supposed to reply immediately with 
a \textsc{Heartbeat\_Ack} message.

Non receiving consecutively, e.g., three \textsc{Heartbeat\_Ack} 
messages from a given monitor leads to its removal from the topology. 
The central server then begins a maintenance (or reorganization) 
phase of the topology.  We estimate that the system can still work 
while there are at least two working monitors.

The second aspect we would like to tackle in the near future is load 
balancing between monitors.  Each traceroute@home monitor focuses on 
its own part of the destination list, as described in 
Sec.~\ref{impl.global}.  However, a problem arises on the step size 
by which to advance the window.  A manually tunable step size does 
not eliminate blocking situations in which a monitor is waiting for 
the prior monitor to terminate its window before it can do any 
further useful work.   Some monitors may potentially wait a long time 
before receiving the needed information (see 
Table~\ref{deployment.success.time}).  This might happen because some 
monitors are slower (as they are more heavily loaded) than others.

We plan to develop, in our future version, a way to balance 
the load between monitors.  This will imply that the sliding 
window size will differ from one monitor to another.  Some monitors 
will work harder while others will maintain a low probing rate.

Currently, a person who controls a traceroute@home system might use 
it in a malicious way in order perform DDoS attacks. Nothing is done 
to prevent this misuse of our tool. However, the centralized solution 
also has the opportunity to improve the security in traceroute@home.

The new version of traceroute@home should also allow communicating 
entities (monitors and servers) to mutually authenticate themselves 
through cryptographic level \dfn{authentication}. The next version 
must also prevent third parties from eavesdropping network 
communications, i.e., guarantee their \dfn{confidentiality}. It is 
unfortunately impossible to fully prevent eavesdropping from 
administrators of machines running our programs.  In addition to 
that, the next version of traceroute@home must be able to guarantee 
the \dfn{integrity} of the results.  We should be able to identify 
tampering if and when it happens.  While it is not reasonable to 
expect that we will be able to detect subtle modifications, we shall 
reject absurd results and stop accepting input from those trying to 
submit them.

We are likely to extensively use cryptographic means. They have all 
the features we need and good  tools are already available. We should 
use a \dfn{public key infrastructure} (PKI) to create and manage 
certificates for both clients and servers. Communicating entities 
will thus have the ability to  easily verify their peer's identity 
before proceeding.

Furthermore, in addition to the Doubletree prototype robustness 
increase, this centralized infrastructure opens interesting 
perspectives, in particular for the development of a general network 
monitoring tool, along the lines suggested by COMNI 
Workshop~\cite{comni} in which we were active.  We can imagine an 
extension to our tracerouting tool in order to provide additional 
measurement services that can be used for network monitoring.  This 
could differ from Scriptroute~\cite{scriptroute} as the monitors have 
the opportunity to cooperate.

Another interesting future undertaking would be to make our 
traceroute@home prototype IPv6 networks aware, allowing thus the use 
of Doubletree in IPv6 networks. Currently, the prototype can only 
probe IPv4 networks. In the near future, we would like to increase 
its capabilities to IPv6 networks.  We believe that the current 
version can be easily extended in order to support IPv6.  The main 
work should be done in the JSocket Wrench library, to handle IPv6 
messages and sockets.  Note that standard IPv6 traceroute, such as 
scamper~\cite{scamper}, or more complex tools, such as 
\dfn{atlas}~\cite{atlas}, already exist.

\subsection{Measurement Infrastructure}\label{further.infra}
%%%%%%%%%%%%%%%%%%%%%%%%%%%%%%%%%%%%%%%
To test our prototype, we choose the PlanetLab infrastructure because 
it offers an easy access to a relatively large quantity of nodes. 
However, despite this apparent simplicity, we encountered several 
difficulties, as mentioned in Sec.~\ref{deployment.prob}.

In a certain sense, these problems were expected as PlanetLab is a 
\textit{testbed} oriented towards overlays and peer-to-peer networks. 
It is not an infrastructure entirely dedicated to network measurement 
or the deployment of measurement tools.  Such tools have to share 
resources (CPU, memory, network access) with all the users, to 
strongly limit their use of disk space, cannot control node
management. 

Consequently and inspired by the COMNI workshop~\cite{comni}, we 
believe it is time to think about an infrastructure entirely 
dedicated to network measurements and network monitoring. This 
infrastructure should allow us to go beyond the experimental
environment of PlanetLab.

In the fashion of PlanetLab, the nodes composing this infrastructure 
should be numerous and geographically diverse.  Further, this 
infrastructure should be carefully engineered in order to avoid 
attacks from the outside world and to avoid abuse from users.

\section{Conclusion}\label{conclusion}
%%%%%%%%%%%%%%%%%%%%
In this report, we described our Java implementation of an efficient 
and cooperative topology discovery algorithm, Doubletree.  We 
implemented the algorithm in a tool we call traceroute@home. 
traceroute@home is freely available and easy to extend.
 
We first discussed the global functioning of the system and, next, we 
introduced the internal architecture of a traceroute@home monitor.  
We also explained the message framework proposed for our prototype.  
This message framework is easy to extend for further improvements.

In order to test our implementation, we deployed our prototype on a
few PlanetLab nodes and evaluated its performance.

We finally identified some weaknesses in our prototype and proposed
several ideas for further development.  We also introduce a discussion
about the opportunity of developing a networking measurement
infrastructure. 

%\section*{Acknowledgements}\label{ack}
%%%%%%%%%%%%%%%%%%%%%%%%%%
%Mr. Donnet's work was partially supported by a SATIN European 
%Doctoral Research Foundation grant and by an internship at CAIDA.

\bibliographystyle{IEEE}
%\bibliography{Prototype}
\bibliography{Bibliography}

\end{document}